\documentclass[aps,prl,twocolumn,groupedaddress,showkeys]{revtex4-1}

\usepackage{graphicx,epstopdf}
\usepackage{subfig}
\usepackage{amssymb,amsfonts,amsmath}

\usepackage[iso]{datetime}

\usepackage{color}
\usepackage[colorlinks=true,linkcolor=blue,urlcolor=blue,citecolor=blue]%
	{hyperref}
\newcommand{\hbIdea}[1]{}

\newcommand{\hbAverage}[1]{\overline{#1}}

\newcommand{\hbie}{i.e., }

\newcommand{\hbQuote}[1]{{\small \textsf{``#1''}}}

\newcommand{\reffig}[1]{Fig.~\ref{#1}}

\newcommand{\reftbl}[1]{Table~\ref{#1}}

\newcommand{\refcite}[1]{ref~\cite{#1}}

\begin{document}

\title{
	Compatibility of Mating Preferences
}

\author{Haluk O. Bingol}
\affiliation{Bogazici University, Istanbul, Turkey}

\author{Omer Basar}
\affiliation{Bogazici University, Istanbul, Turkey}


\begin{abstract}
	Human mating is a complex phenomenon.
	Although men and women have different preferences in mate selection, 
	there should be compatibility in these preferences
	since human mating requires agreement of both parties. 
	We investigate 
	how compatible the mating preferences of men and women are in a given property
	such as age, height, education and income.
	We use dataset of a large online dating site  ($N = 44,255$ users).
	(i)~Our findings are based on 
	the ``actual behavior'' of users trying to find a date online,
	rather than questions about a ``hypothetical'' partner as in surveys.
	(ii)~We confirm that women and men have different mating preferences.
	Women prefer 
	taller and 
	older men with 
	better education and 
	higher income then themselves.
	Men prefer just the opposite.
	(iii)~Our findings indicate that these differences complement each other.
	(iv)~Highest compatibility is observed in income with $95~\%$.
	This might be an indication that
	income is in the process of becoming more important than other properties, 
	including age, in our modern society.
	(v)~An evolutionary model is developed which produces similar results.
\end{abstract}

%


\maketitle

\section*{Introduction}
\label{sec:Introduction}

\subsection*{Mating and Parental Investment}

\hbIdea{mating and parental investment theory} 
Mating is important for evolution.
In many species, it has been observed that 
males and females have different strategies in mate 
selection.
An evolutionary theory 
to explain differences in mating strategies is 
Trivers' 
\emph{parental investment theory}~\cite{%
	Trivers1972}.
He carefully defines \emph{parental investment} as
	\hbQuote{any investment by the parent in an individual offspring 
		that increases the offspring's chance of surviving 
		(and hence reproductive success)
		at the cost of the parent's ability to invest in other offspring}~\cite{%
	Trivers1972}.
Therefore, 
evolution calls for parental support,
since offsprings with more parental support have better chance to reproduce.

\hbIdea{mating strategies} 
Parental investment is quite uneven between male and female
in many species~\cite{%
	Trivers1972}.
Therefore, both genders evolutionarily developed mating strategies, 
which are clearly different~\cite{%
	Buss1989,
	Buss1993,
	Buss2006,
	Buss2003Book}.
\hbIdea{female strategy} 
Human female makes mandatory high investment in offspring compared to male,
if one considers nine months of gestation, childbirth, lactation, nurturing.
Therefore, 
she looks for a supporting male in her mate selection.
She prefers a male who 
not only has the resources to support her 
but 
also willing to commit these resources to her.
This explains female preference for \emph{long-term} commitment.
\hbIdea{woman strategy} 
On the other hand, 
human male can prioritize quantity.
Relative to the female, 
he is reluctant to engage in long term commitment, 
partly due to \emph{parental uncertainty},
that is, 
he cannot be hundred percent sure that the child carries his genes.
He has a tendency for \emph{short-term} relations,
which increases his chances to reproduce offsprings.
This quality versus quantity trade-off creates a conflict 
that has to be resolved.
Females, 
who invest more in offspring,
should be more choosy selecting a mate
(\emph{intersexual attractio}n)
and
males,
who invest less,
should compete with other males to access the opposite sex
(\emph{intrasexual competition})~\cite{%
	Trivers1972}.

\subsection*{Properties in Mate Selection}

\hbIdea{mating properties} 
Properties 
that increase the chance of mating 
become crucial in this respect~\cite{%
	Silverman1992,  
	Trivers1972,
	Buss1989,
	Buss1993,
	Buss2006,
	Kenrick1992, 
	Gillis1980, 
	Pawlowski2000Nature,  
	Swami2008,  
	Buss2003Book}.  
\hbIdea{age} 
In terms of evolution,
(i)~\emph{fertility},
\hbie~immediate probability of conception,
and
(ii)~\emph{reproductive value},
\hbie~future reproductive potential,
are the top two properties for both 
genders~\cite{%
	Buss2006}.
In many species, younger and healthier members are more likely to have these properties,
which makes them more attractive as mates.
Human males are particularly tuned to these properties in mate selection.
However, because of the immense cost of reproducing, 
human females are attentive to other properties as well,
namely
(i)~the ability to gather resources and 
(ii)~the willingness to commit these resources to her offspring.
Human societies are heavily hierarchical 
with those at the top typically having much more power and access to material resources.
It takes time for a typical man to rise up in the hierarchy.
Therefore, older men are more likely to have the properties desired by women than their younger counterparts.
\hbIdea{physical} 
In addition, 
men that are physically stronger or otherwise advantageous (e.g., taller) 
will be better able to protect a dependent (pregnant or nursing) mate and
her vulnerable offspring.
Hence, physically masculine men should be preferred by women~\cite{%
	Pawlowski2000Nature}. 
\hbIdea{expect} 
In sum, 
we expect that youth in women, 
and older age and masculinity in men are properties 
that complement each other in mate selection.

Empirical evidence supports these 
deductions~\cite{%
	Buss2003Book,
	Schmitt2008,
	Silverman1992, 
	Buss1989,
	Kenrick1992, 
	Gillis1980, 
	Pawlowski2000Nature,  
	Swami2008}.  
In couples,
\hbIdea{age} 
the man is typically older~\cite{%
	Kenrick1992}, and
\hbIdea{physical} 
taller~\cite{%
	Gillis1980,
	Pawlowski2000Nature,
	Swami2008}
than the woman.
This is a universal pattern across cultures~\cite{%
	Buss1989,
	Schmitt2008}.

\hbIdea{Compatibility of Preferences} 
Since mating requires agreement of both parties, 
although men and women have different preferences in mate selection, 
there should be compatibility in these preferences. 
The question that guides the present research is:
How compatible are the mating preferences of two genders regarding a given property?

\section*{Method}

We use the data obtained from an online dating site. 
First, we carefully define ``mating'' in our data set. 
	(See 
	\hyperref[sec:DataSet]{Data Set}
	and
	\hyperref[sec:DefinitionOfMating]{Definition of Mating}
	in the Appendix.) 
Then we aggregate the properties of partners that an individual selects as mates.
Finally we search for patterns in the properties for mating behavior.

\subsection*{Definitions}

\subsubsection*{Properties of the Mate}

\hbIdea{properties of the mate} 
Once we have identified the partners,
we investigate the properties of the mate.
As expected, 
user $i$ becomes partner with various others in time.
Each partner of $i$ may have a different value for property $p$.
The average of the properties of the partners of $i$ 
is given as
\[
	\hbAverage{p_{i}} = \frac{1}{|C_{i}|}  \sum_{j \in C_{i}} p_{j}
\]
where 
$C_{i}$ is the set of users that $i$ partnered with,
and 
$p_{j}$ denotes the property $p$ as it is defined in user $j$'s profile.
We interpret this as follows: 
User $i$ has a tendency to select partners having value of 
$\hbAverage{p_{i}}$ in property $p$.
Hence, 
we call $\hbAverage{p_{i}}$ as the \emph{preferred value} for $i$.

\subsubsection*{Preferred Difference}

Instead of using the preferred value directly,
we compare one's own value
to the preferred value that one looks for in his partners.
The \emph{preferred difference} of $i$, 
in property $p$, 
is defined as
\[
	\Delta p_{i} =  \hbAverage{p_{i}} - p_{i}.
\]
Note that $\Delta p_{i}$ can be positive or negative. 
If $\Delta p_{i}$ is around $0$
then the user prefers partners with similar properties with her, 
i.e. homophily~\cite{%
	McPherson2001,
	Centola2011Science}.
If user $i$ has a tendency to select partners that are taller than herself,
then $\Delta p_{i}$ in height would be positive;
otherwise it would be negative.

\subsubsection*{Preferred Difference Distributions}

\hbIdea{property distribution} 
We can extend these concepts from individual $i$ to a group of people.
Then, frequency of people with the same preferred difference 
makes a probability distribution,
which we call \emph{preferred difference distribution}.
Having 
all women as one group, and 
all men as another group,
we obtain two preferred difference distributions
$f(x)$ and $m(x)$ of 
females and males,
respectively.

\section*{Results}

We investigate four properties, 
namely, 
age, height, income and education.
As already discussed in the
\hyperref[sec:Introduction]{Introduction} section,
age and masculinity are important properties in mating for all species.
Height is considered as an indication of masculinity.
The other two properties, income and education, are unique to humans.

\hbIdea{findings} 
The statistical parameters of the preferred difference distributions in 
age, height, education, and income are given in
\reftbl{tbl:statistics}.
Columns 
$\mu_{m}$,  $\mu_{f}$, 
and
$\sigma_{m}$, $\sigma_{f}$ are the
averages and standard deviations of men and women, respectively.
We first focus on the averages,
and leave the discussion of 
the distributions of the preferred differences,
and 
their compatibility $\rho$ for later.

\begin{table*}[!htbp] 
\centering
	\caption{
		{\bf 
			Comparison of male and woman distributions 
		}
	}
	\begin{tabular*}{\hsize}{@{\extracolsep{\fill}}|l|rr|rr|c|}
		\hline
		{\bf Property}%
		&\multicolumn{2}{c|}{{\bf Averages}}
		&\multicolumn{2}{c|}{{\bf Standard Deviations}}
		&{\bf Compatibility}\\
		~%
		&$\mu_{m}$
		&$\mu_{f}$
		&$\sigma_{m}$
		&$\sigma_{f}$
		&$\rho$\\
		\hline
		Income (bin)
		&-0.93
		&0.99
		&1.28
		&1.32
		&0.95\\
		Age (year)			
		&-2.90
		&2.74
		&5.06
		&5.23
		&0.94\\
		Education (bin)
		&-0.36
		&0.34
		&1.35
		&1.40
		&0.92\\
		Height (cm)		
		&-11.12 
		&11.37 
		&6.76
		&7.09
		&0.90\\
		\hline
	\end{tabular*}
	\label{tbl:statistics}
\end{table*}

\subsection*{Averages of Preferred Differences}

In all four properties in \reftbl{tbl:statistics},
there is a distinct pattern: 
The averaged preferred differences 
for men, $\mu_{m}$, are all negative 
and
those of women are all positive. 
This observation indicates that in all four properties, 
regardless of the metric that is used to measure the property,
men prefer women with lower scores
and women prefer men with higher scores, 
compared to themselves.

\subsubsection*{Height}

\hbIdea{difference in partner selection} 
Our findings on preferred difference in height, given in 
\reffig{fig:propertyDifferenceHeight}, 
do not agree with the similarity hypothesis in mate selection.
Only a minority of men and women are accumulated around the zero point, 
which is the point of no preferred difference (i.e., maximum similarity).
Instead, 
the averages in \reftbl{tbl:statistics}
agree with the male-taller norm.
On average, 
men prefer women 
$11.12$\,cm shorter.
Similarly, 
on average
women prefer men 
$11.37$\,cm taller.
The distributions in 
\reffig{fig:propertyDifferenceHeight} 
clearly show that the majority of men prefer shorter women and, 
in complementary fashion,
the majority of women prefer taller men.  Thus, online daters are attracted to others who complement their height preferences. 
These findings replicate previous work on
the male-taller norm~\cite{%
	Gillis1980,
	Swami2008
	}.

\newcommand{\myFigWidthOne}{1\columnwidth}
\begin{figure*}[th] 
\centering
	\subfloat[Height]{%
		\includegraphics[width=\myFigWidthOne]%
			{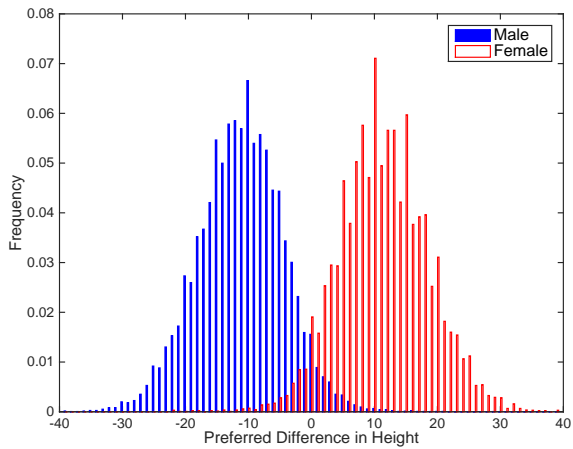}
		\label{fig:propertyDifferenceHeight}%
	}
	\subfloat[Age]{
		\includegraphics[width=\myFigWidthOne]%
			{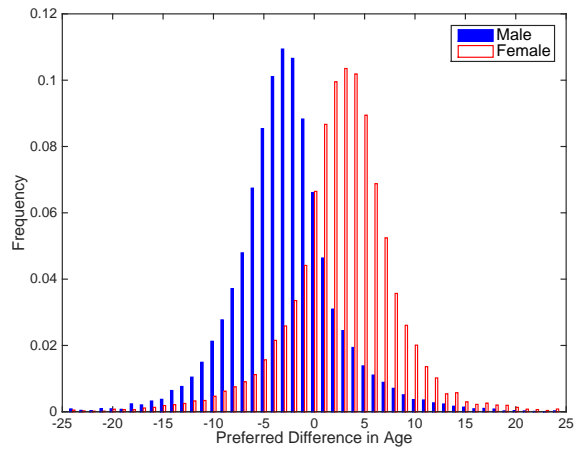}
		\label{fig:propertyDifferenceAge}
	}\\
	\subfloat[Education]{
		\includegraphics[width=\myFigWidthOne]%
			{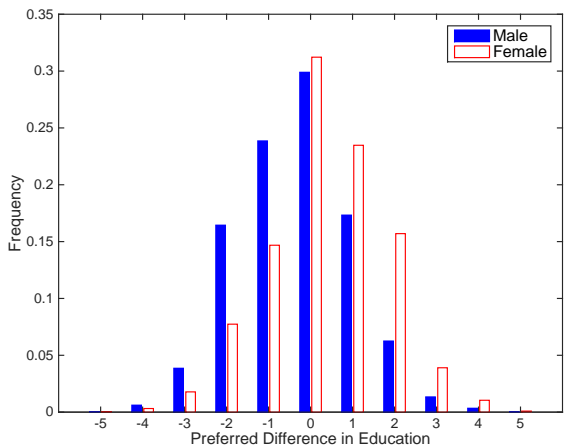}
		\label{fig:propertyDifferenceEducation}
	}
	\subfloat[Income]{
		\includegraphics[width=\myFigWidthOne]%
			{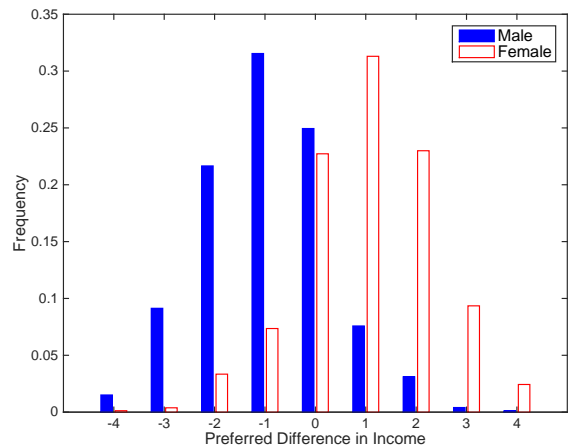}
		\label{fig:propertyDifferenceSalary}
	}
	\caption{
		(Color online)
		{\bf 
		Preferred difference distributions. 
		}
	}
	\label{fig:propertyDifference}
\end{figure*}

\subsubsection*{Age}

\hbIdea{age} 
According to evolutionary theories,
we expect to see a greater prevalence of 
younger woman-older man partners 
than alternative age couplings.
Our findings reveal such a pattern in mate preferences in online daters.
Examination of age preference distributions, given in 
\reffig{fig:propertyDifferenceAge}, 
indicates that the majority of men prefer younger women 
whereas the majority of women prefer older men.
On average, 
men mate with women $2.90$ years younger than themselves,
and
women mate with men $2.74$ years older than themselves.
Our findings are in agreement with 
Buss~\cite{Buss1989},
which reports that
men prefer 2.66 years younger, 
women prefer 3.42 years older mate than themselves.

\subsubsection*{Income and Education}

We observe similar patterns in income and education, as well.
Namely,
men prefer negative and women prefer positive differences.
Here the numbers cannot be compared with other works directly
since 
users are asked to select one bin out of many bins, 
which are organized in a consistent but arbitrary way~\cite{%
	Bingol2012PartnerArxiv}. 
They are consistent in the sense that the larger the bin number,
the more educated or higher income.
The bins in education are arranged according to the years spent in school 
such as
graduate of primary school, or
of college.
The bins in the income field represent monthly income such as 
bin-2:~$500 < x < 1,000$,
bin-3:~$1,000 < x < 2,000$.



\subsection*{Distribution of Preferred Differences}

Note that the absolute values of the average preferred difference of men and women, 
as well as the standard deviations of preferred differences, 
are close to each other in \reftbl{tbl:statistics}.
We aimed to examine whether this was coincidental or substantive. 

For this purpose, 
we relied on the distributions of preferred differences 
in height, age, education, and income that are given in 
\reffig{fig:propertyDifference}.
The bell-shaped curves of male and female distributions 
resemble to each other. 
Female curves are right-shifted
while 
male curves are left-shifted, 
with respect to the $y$-axis. 

\hbIdea{compatible} 
How do we compare these curves?
In order to get a better understanding,
consider a simplified example given in 
\reffig{fig:propertyDifferenceDummy}.
Note that
women that prefer $\Delta p = x$
match with 
men that prefers $\Delta p = -x$.
Therefore,
we should not directly compare the distribution $f(x)$ of women
with $m(x)$ of men.
We should compare $f(x)$ with $m(-x)$,
that is,
the symmetric graph with respect to the $y$-axis.
We make the reasonable assumption that 
there are equal number of men and women.
Then,
$\min\{ {f(x), m(-x)} \}$ of the women who prefer $\Delta p = x$ are matched.
Thus, 
the \emph{compatibility} of two distributions can be measured by means of the ratio of matched women given as
\[
	\rho = \sum_{x} \min\{ {f(x), m(-x)} \}
\]
where summation is taken over all possible values of $x$.
Note that $0 \le \rho \le 1$ 
where 
$\rho = 1$ means women and men are perfectly compatible.
This is a well-defined metric since 
the ratio of matched women is equal to that of men.

\newcommand{\myFigWidthTwo}{\columnwidth} 
\begin{figure}[th] 
\centering
	\includegraphics[width=\myFigWidthTwo]%
		{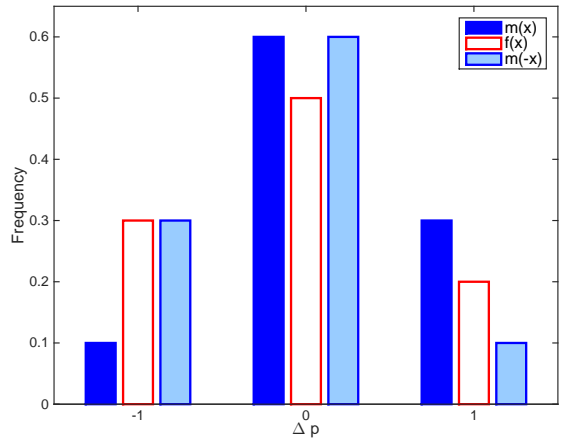}%
	\caption{
		(Color online)
		{\bf Distribution of difference in 
		a dummy property $p$.}
		We assume that 
		$\Delta p$ takes values of $-1$, $0$, and $1$. 
		We also assume that 
		woman and male populations are of the same size.
		(i)~$50~\%$ of women and 
		$60~\%$ of men prefer no differences in $p$.
		Hence $50~\%$ matches for $\Delta p = 0$.
		(ii)~$20~\%$ of women who prefer difference of $\Delta p = 1$ 
		are to match with
		$10~\%$ of men who prefer differences of $\Delta p = -1$.
		Only $10~\%$ matches for $\Delta p = 1$.
		(iii)~$30~\%$ of women who prefer difference of $\Delta p = -1$ 
		are exactly match with
		$30~\%$ of men who prefer differences of $\Delta p = 1$.
		That is, $30~\%$ matches for $\Delta p = 1$.
		In total $90~\%$ of women are matched.
		Hence male female compatibility is $\rho = 0.90$.
	}%
	\label{fig:propertyDifferenceDummy}%
\end{figure}


\section*{Model}

The properties are listed in descending order of compatibility in \reftbl{tbl:statistics}.
Height is the property with the lowest compatibility.
Even for this property,
$90~\%$ of the population can find a partner who satisfies one's preferences.
What kind of dynamics can lead the system to such a high compatibility?

A simple evolutionary model can explain high compatibility values.
	(See 
	\hyperref[sec:DetailsOfTheModel]{Details of the Model} 
	in the Appendix.) 
As an abstraction, 
we consider agents with one property only, 
such as age.
It starts with male and female population with diversified values in the property, 
hence low compatibility.
As the system evolves, 
the compatibility increases as seen in \reffig{fig:model}.

\newcommand{\myFigWidthThree}{\columnwidth}%
\begin{figure}[th] 
\centering
	\includegraphics[width=\myFigWidthThree]%
		{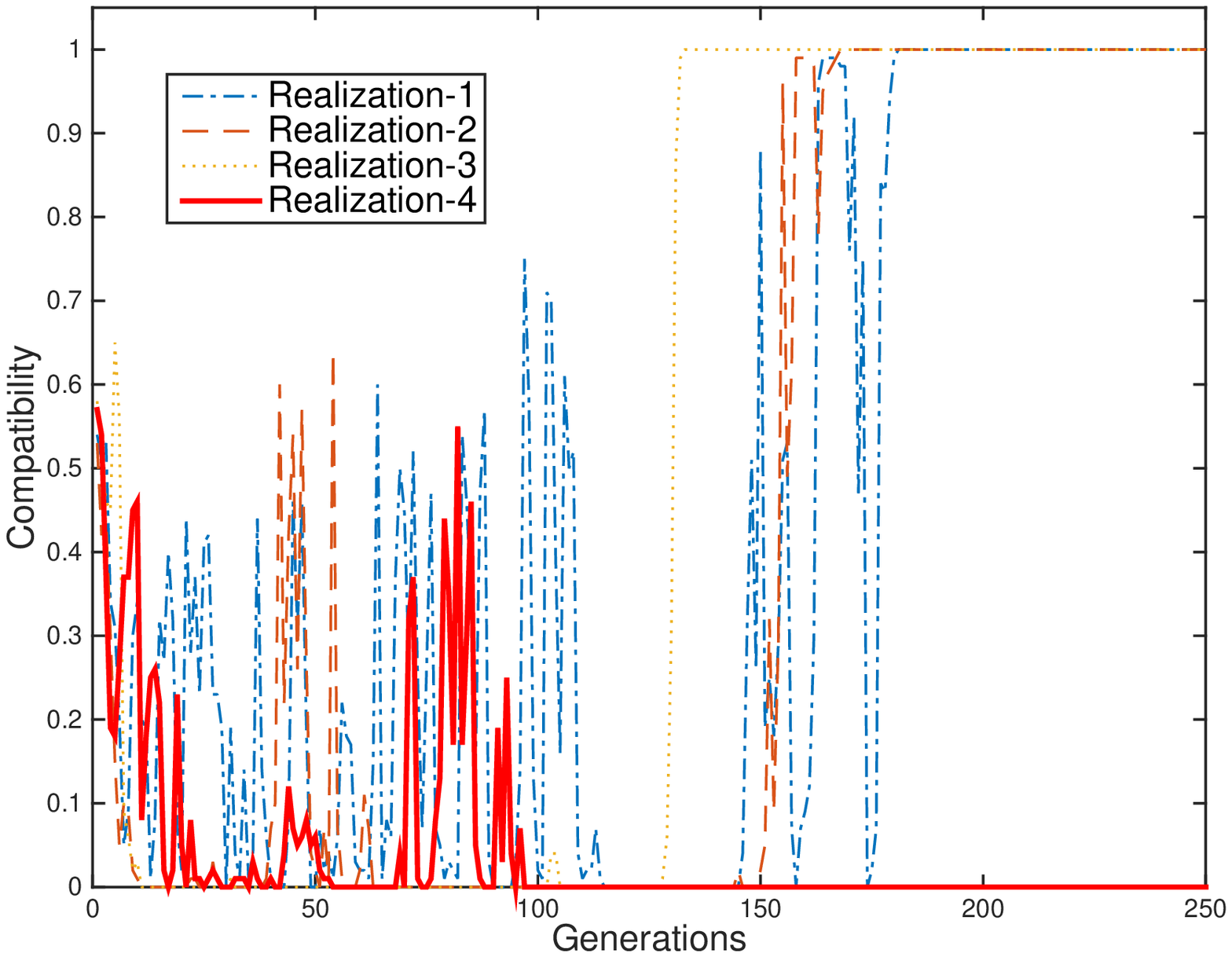}%
	\caption{
		(Color online)
		{\bf Sample realizations of the model.}
		Many realizations survive and reach to the perfect compatibility of 1.
		Only a few become extinct.
		As a representative of extinction,
		Realization-4 is included to the plot.
		($N = 100$, $R = 9$, $M = 20,000$).
	}%
	\label{fig:model}%
\end{figure}

The increase  of compatibility is due to the decrease of genetic variation.
Simulations of the model reveal that 
genetic variation is reduced from generation to generation.
Agents, 
that cannot find mates, 
are eliminated from the system.
Agents, 
whose genotype fits the population,
survive.
Hence the compatibility of the population increases.
It is observed that 
the system converges to 
the perfect compatibility most of the time.
Convergence is quite fast. 
For $N = 100$,
no more than 200 generations are usually sufficient.
Of course,
there are some realizations that become extinct 
but they are rare.
Interestingly,
in many realizations 
genotype pool reduces to
one female and one male genotype.


\section*{Discussion and Conclusions}

Men and women behave differently~\cite{%
	Liljeros2001Nature,
	CelaConde2009PNAS,
	Hyde2009PNAS,
	Szell2013SR,
	Blanch2015}.
Our findings show that  
the virtual world of online dating is another manifestation of gender differences.

\subsection*{Opposite Preferences in Scores}

\hbIdea{discussion} 

While women prefer men with higher scores in every property 
that we have investigated, 
men show just the opposite pattern. 
This supports findings from previous studies but with the added benefits of
(i)~focusing on the actual mating behaviors of real people, 
and 
(ii)~drawing on such a large sample of $N = 44,255$.

\subsection*{Compatibility of Preferences}

Interestingly, the preferences of men and women match each other so that 
the number of dissatisfied seekers is minimized.
This can be explained by evolution. 
Individuals with preferences 
that do not match 
what is available in the current population 
have restricted mating opportunities and, consequently, lower reproductive success.
The proposed model shows that, 
as it evolves, 
a population becomes more homogeneous in terms of variation in the preferences.
The genomes that fit the population are selected more and eventually dominate the population.

\subsection*{Ranking of Properties}

The mating properties should be ranked in terms of importance.
We expected that 
the system should tune itself to maximize compatibility first in age and then in height. 
Interestingly,
it is observed that income is the property with
the highest compatibility of $95~\%$.

One possible explanation might be the change in society across time.  
Income may have taken precedence over other properties, such as age, 
in today's society.  
High income has great utility for solving evolutionary problems, 
if one considers the impact of accessing medical care for health, 
and also medical support for reproduction.
This may be the reason 
why both genders in contemporary society have become especially attuned to income 
and consequently learned to adapt their preferences for maximum compatibility with the opposite gender. 

Age may have been a much more reliable cue to a man's ability to
obtain status, wealth, and resources in the ancestral environment than in today's society.  
However, age is still an important cue for female fertility 
and thus follows income as the property with the second highest compatibility level $94~\%$. 
Clearly, 
$1~\%$ difference between income and age makes these suggestions highly speculative
and calls for further research.

In our modern society,
we may speculate that ability to protect a woman partner and the offspring physically,
which may be facilitated by height,
loses importance, 
leading to height being the property with the lowest level of compatibility in the current sample.

Education, 
which can be related to social position and correlated with income,
gets the third place as height loses its importance.

\subsection*{Actual versus Hypothetical Preferences}

As a final but important point,
we would like to stress that our findings are not based on surveys, 
in which
one answers questions about a ``hypothetical'' partner, 
one wishes to have,
as in many other works such as \refcite{%
	Buss1989}.
Our findings are based on 
the ``actual behavior'' of users trying to find a date online.
Our data tracks ``actual'' partners,
that is,
people who have mutually agreed to ``mate'' 
as far as one can trace in an online dating site.

In addition, as in many surveys,
we do have data about ``hypothetical'' mate preferences, too. 
Users specify what properties, 
such as age and height, 
they look for in their potential partners.
However, such self-reported preferences
are not consistent with actual behavior.
For example,
someone might claim that he prefers women taller than 170~cm but
show no hesitation to partner with a woman who is 160~cm.
Such inconsistencies are not apparent 
if research focuses only on questionnaire responses.
In this respect our behavioral data deserves special attention.

\section*{Appendix}

\subsection*{Data Set}
\label{sec:DataSet}
	
\hbIdea{online dating data} 
We investigate the data of a large online dating site 
for compatibility of mating preferences~\cite{%
	Bingol2012PartnerArxiv}.
There are 4,500,000 registered users in total.
More than 3,000 new users register daily. 
Users stay in the system for three months on average. 
Many of those who leave come back later;
sometimes as a new user.
The daily activity is also quite large in volume with 
50,000 user logins,
500,000 messages, 
and 
20,000 votes (of other's profiles).

\subsubsection*{Privacy and Data Availability}

Privacy is the most important issue for such an investigation.
In this study, 
all data gathering and data processing is done at the company site.
No data left the company. 
Only statistical data such as the histogram data visualized in 
\reffig{fig:propertyDifference} is shared with us.
This histogram data is available at \cite{bingolXXXXData}.

\subsubsection*{Other Issues}

\hbIdea{warnings} 
One needs to be careful on a number of issues in a study like the present one.
(i)~The user defines his properties in the profile.
Hence, user properties may be misleading.
On the other hand, 
stretching the properties too far would not be a good strategy 
since unfaithful declaration, 
such as claiming to be slim while actually being obese, 
would be an obstacle to further the relationship
when the time comes to meet face-to-face~\cite{%
	Norcie2013LNCS,%
	Ellison2013}.
So we assume that users are close to what they claim to be.
(ii)~One has to keep in mind that 
the findings could be culture dependent.
(iii)~We focus on heterosexual relations only. 
Wholly other dynamics might be in motion in non-heterosexual relationships.

\subsection*{Definition of Mating}
\label{sec:DefinitionOfMating}

\hbIdea{online dating} 
A typical online dating system enables its user to 
find a partner that best matches one's preferences.
Each user defines his or her user profile.
An initiator, predominantly man, 
selects a potential partner by examining her profile and 
sends her a message.
If there is a positive response from the receiver, 
then more messages are exchanged
which may eventually lead to a face-to-face meeting.

\hbIdea{partner criteria} 
When do we say that a man and a woman are mating?
Online dating sites contain abundant information in the virtual domain (i.e., profiles), 
but there is usually no information whether 
the man and the woman are actually mating in the physical world.
Any action in an online dating site is clearly an attempt for mating
but is this sufficient to be considered mating?
For example,
just sending a message,
getting a message in response, 
or even exchanging a series of messages should not qualify as mating 
since the nature of just seeking a partner online already involves these activities.

\subsubsection*{Virtual Gifts}

Therefore we select  
the most restricted criteria of mutual interest 
that is available in our data set,
which is based on virtual gifts~\cite{%
	Bingol2012PartnerArxiv}.
Receiving a \emph{virtual gift}, 
which is usually a picture of a flower, 
is considered  a ``value'' in this virtual society.
We have even observed that some users sent virtual gifts to themselves.
This value is probably due to a number of reasons:
(i)~The virtual gifts one receives is visible to all.
(ii)~They are not free, 
\hbie\ one has to purchase virtual gifts with actual money.
(iii)~Only qualified users can send virtual gifts.
Since unpaid male members are not qualified to send gifts, 
being able to send gifts may be considered an indication of wealth.

Considering virtual gifts drastically reduces the subject size.
Within $4,500,000$ registered users;
there are $276,210$ men and 
$483,963$ women 
that are qualified to send virtual gifts in the system. 
Among those, only 
$29,274$ men and 
$14,981$ women,
in total $N = 44,253$, 
users reciprocally exchange virtual gifts.
Hence we define a pair as (\emph{mating}) \emph{partners} if
they have exchanged (i.e., send and receive) at least one gift.

\subsection*{Details of the Model}
\label{sec:DetailsOfTheModel}

\subsubsection*{Agents}

Agent $i$ represented by a 4-tuple $(g_{i}, p_{i}, p_{i}^{\min}, p_{i}^{\max})$,
called \emph{genotype},
where
\emph{gender} $g_{i}$ is a binary number with 0 being female.
The property takes values in the \emph{value range} of 
$\{ 1, 2, \dotsc, R \}$ 
for some positive integer $R$.
The \emph{property} of the agent is denoted by $p_{i}$.
The values $p_{i}^{\min}$ and $p_{i}^{\max}$ 
represent a range for possible mate.
That is,
the agent ``agrees'' to mate with an agent with $p_{j}$ 
only if
$p_{i}^{\min} \le p_{j} \le p_{i}^{\max}$.
We will get back to mating shortly.

\subsubsection*{Generations}

We have to initialize the very first generation.
The rest of the generations are driven by the system.
The first generation is composed of $N$ female and $N$ male agents.  
We initialize female and male agents differently.
In order to initialize an agent, 
we draw three numbers from a uniform random distribution in the value range 
and order them so that we have
$p_{1} \le p_{2} \le p_{3}$.
Then if the agent is female, 
set its genotype to
$(0, p_{1}, p_{2}, p_{3})$,
otherwise to
$(1, p_{3}, p_{1}, p_{2})$. 
Note that 
we use the minimum of the numbers for her score
since females prefer males that have higher scores than theirs.
Similarly,
males prefer females with lower scores.
That is, our model starts with agents that are agree with the 
``female prefers higher scores, male prefers lower scores'' assumption.

Life span of a generation is defined as 
$M$ meetings in total.
After $M$ meetings, 
the parents are removed and 
the children become the next generation.
Note that the number of girls and the boys are usually different than $N$.
If the population of any gender exceeds $N$,
we randomly eliminate some so that
every generation has no more than $N$ females and $N$ males.
If the population of a gender is less than $N$, 
we do not do anything to increase it to $N$.
Of course,
if the population of one gender becomes zero,
then the system stops reproducing.

\subsubsection*{Mating and Reproduction}

We let agents obtain the chance to reproduce 
by randomly picking a female and a male to \emph{meet}.
A meeting produces a child if
both agents ``agree'' to mate.
Agreement is defined as follows.
Agent $i$ \emph{agrees} to mate with the opposite gender agent $j$
if $j$ is in the mating range of $i$,
that is,
$p_{i}^{\min} \le p_{j} \le p_{i}^{\max}$.
For example, 
male $(1, 5, 2, 5)$ 
accepts 
female $(0, 2, 6, 8)$ 
but the female does not accept the male.

The child is set to be female with probability $0.5$.
We used a very simple inheritance mechanism: 
the daughter gets the properties of the mother and  
the son gets that of the father.
Therefore, the genotype of either the mother or the father is 
preserved in the next generation. 
Since a child gets exactly the same genotype of either parent,
and 
since agents in our initial generation agree with the 
``female prefers higher scores, male prefers lower scores'' rule, 
this rule is preserved in all generations.

\subsubsection*{Genotype Variety}

We obtain the first generation by repeating agent initiation process 
$N$ times for female and 
$N$ times for male.
Hence we expect that the initial generation has quite a variety of genotypes, 
possibly $2N$ different genotypes.
Since there is 
neither mutation 
nor recombination of genotypes,
there is no way that the genotype variety of the system can increase. 
On the contrary, 
it can reduce 
when the last member of a genotype fails to reproduce.

Note that we do not consider mutations in this very simple model.
One may extend the model by introducing mutations.

\section*{Acknowledgments}
This work was partially supported 
by Bogazici University Research Fund, BAP-2008-08A105, 
by the Turkish State Planning Organization (DPT) TAM Project, 2007K120610.
The authors would like to thank 
Adil Saribay for his assistance with psychology literature,
Resit Canbeyli for constructive comments, 
and
Melih Barsbey for careful review of the draft.
H.O.B. would like to thank 
to the people of COST Actions MP0801 
and 
especially to Marcel Ausloos 
for stimulative discussions.





\end{document}